\documentclass[10pt,conference,letterpaper,doublecolumn]{IEEEtran}

\usepackage[dvips]{color}
\usepackage{epsf}
\usepackage{times}
\usepackage{epsfig}
\usepackage{graphicx}
\usepackage{url}
\usepackage{amsmath}
\usepackage{amssymb}
\usepackage{amsxtra}
\usepackage{algorithmic}
\usepackage{algorithm}
\usepackage{cite}

\newtheorem{theorem}{\bf Theorem}

\newtheorem{definition}{\bf Definition}

\begin{document}

\title{Dynamic Interference Minimization Routing Game for On-Demand Cognitive Pilot Channel\vspace{-3mm}}

\IEEEoverridecommandlockouts

\author{Quanyan~Zhu$^{\dag}$,~
Zhou~Yuan$^{\ddag}$,~Ju~Bin~Song$^{\S}$,~Zhu Han$^{\ddag}$,~and$~$Tamer~Ba\c{s}ar$^{\dag}$
\thanks{$^\dag$Q. Zhu and T. Ba\c{s}ar are with the Coordinated Science Laboratory and the Department
of Electrical and Computer Engineering, University of Illinois at Urbana Champaign,
IL,  USA, 61801 e-mail:\{zhu31, basar1\}@illinois.edu. 
$^\ddag$Z. Yuan and Z. Han are with the Department of Electrical and Computer Engineering, University of Houston, Houston, TX, USA, e-mail: \{zyuan, zhan2\}@mail.uh.edu).
$^{\S}$J. B. Song is with the Department of Electronics and Radio Engineering, Kyung Hee Univiversity, Yongin, S. Korea, e-mail: jsong@khu.ac.kr.
Research supported by Korea Science and Engineering Foundation (20090075107), NSF CNS-0953377, CNS-0905556, and CNS-0910461.}
}

\maketitle

\begin{abstract}
In this paper, we introduce a distributed dynamic routing algorithm for secondary users (SUs) to minimize
their interference with the primary users (PUs) in multi-hop cognitive radio
(CR) networks.  We use the medial axis with a relaxation factor as
a reference path which is contingent on the states of the PUs.  Along the axis, we construct a hierarchical structure for
multiple sources to reach cognitive pilot channel (CPC) base
stations. 
We use a temporal and spatial dynamic non-cooperative game to model the interactions among  SUs as well as their influences from PUs in the multi-hop structure of the network. A multi-stage fictitious play learning is used for distributed routing in multi-hop CR networks. We obtain a set of mixed (behavioral) Nash equilibrium strategies
of the dynamic game in closed form
by  backward induction. The proposed
algorithm minimizes the overall interference and the average packet delay along the routing path from SU nodes to CPC base stations in an optimal and distributed manner. 

\end{abstract}


\section{Introduction}
The primary users (PUs) directly affect the spectrum opportunities
available for the secondary users (SUs). As a consequence, it results in the time-varying
wireless channel conditions and the dynamic network topology in multi-hop
cognitive radio (CR) networks \cite{Han2008, Hossain2009,
Schaar2009}. 
Recently, the cognitive pilot channel (CPC) has been suggested to
provide frequency and geographical information to SUs, assisting
them in sensing and accessing the spectrum 
 \cite{Evans2005,Cordier2009, Guipponi2007}. As a result, the SUs can improve their performance
and avoid scanning the entire spectrum to identify the spectrum
holes and available PUs. The on-demand CPC transmits the information
only at the request of a SU terminal. The on-demand CPC
including both an uplink and a downlink channel enables a wider
range of CPC-based applications in addition to the retrieval of the
information about operators, radio access technologies, and
frequency lists. To establish an effective CPC network, a multi-hop
distributed CR network scheme is required \cite{Schaar2009,
Guipponi2007}. The main existing work in this area has focused on the
information contents that CPCs can convey, as well as on
the implementation aspects of the channels. To the best of our knowledge, no work has been done to
investigate an optimal dynamic routing scheme for the on-demand
CPC  to route requests and deliver the CPC information.

Due to the dynamic interception of frequency channels by
PUs, SU networks should dynamically update their routing paths.
Although a few distributed routing algorithms have been suggested for
multi-hop CR networks \cite{Hou2008, Medley2009}, it is imperative to study
an effective, dynamic and intelligent routing scheme  in CR
networks that considers the time-varying channel states and minimizes the interference not only over the routing path but also over a long time horizon. 
In \cite{Han2009}, a network formation game algorithm  has
been studied for multi-hop CDMA networks in which wireless users attempt to connect to the base station via other users in the cellular network. The routing scheme is based on a spatial dynamic game in which each user optimizes the multi-stage interference along the path. In this paper, we consider a similar dynamic
routing game framework for the application to multi-hop CR networks. Given the time-varying nature of the PUs, we investigate a spatially and temporally dynamic game framework which takes into account the state variation of the PUs as well as the multi-stage property of the CR networks.

The main contribution of this paper is to provide a distributed and optimal dynamic multi-hop network routing scheme for the on-demand CPCs. We consider a CR network comprised of PUs and SUs. The SUs form multi-hop hierarchal levels to the CPC base stations and their performance is based on the location of PUs and their states. In this work, we use thresholds to separate the SUs  logically  into different layers and allow SUs at each level to play a game against other users by choosing the optimal route. In  simulations, we observe that with the presence of a PU, the SUs deviate from their original routes to avoid collisions with the PU. The proposed algorithm minimizes the interference with PUs and packet delay along the routing path.

The rest of the paper is organized as follows. Section II presents the system model, and the game-theoretical model for the CPC network routing is described in Section III. In Section IV, we analyze the dynamic game and characterize the mixed Nash equilibrium in a recursive form. In addition, we devise an algorithm based on the fictitious play learning for the dynamic multi-hop network routing game. The simulation results are described and analyzed in Section V. Finally, conclusions are drawn in Section VI.

\section{System Model}\label{sec:system}
In  CR networks with the presence of PUs who constantly intercept frequency channels, the multi-hop routing of SUs needs to be made dynamic, distributed and efficient to reduce their total interference with the PUs. We assume that  SU nodes are capable of acquiring the knowledge of the channel conditions and  their neighboring relays. SUs dynamically form routing patterns and allocate transmission channels in the spectrum holes of the CR networks.
Unlike multi-channel multi-radio schemes, which operate on one channel at a time, multi-hop CR networks can switch frequency channels on a per-packet basis.
In this paper, we consider the scenario in which the network relays are CR SU nodes and the data is relayed from source nodes to CPC base stations.
We assume that the data format of the on-demand CPC channel is determined, and consequently, the bandwidth of a frequency channel is fixed for every SUs.

Let $ \mathcal{G} = (\mathcal{N}, \mathcal{E})$ be a topology graph for a multi-hop CR network,
where $\mathcal{N}=\{n_{1}, \ldots, n_{\textit{N}} \}$ is a set of $\textit{N}$ SU nodes including the source SU nodes and relay SU nodes;  and
$\mathcal{E}=\{e_{1}, \ldots,  e_{\textit{E}} \}$ is a set of $\textit{E}$ links connecting the SUs and the relay nodes.
In addition, we let $\mathcal{M}=\{m_{1}, \ldots,  m_K \}$ be a set of $K$ PUs and $\mathcal{K}$ be the set of CPC channels.
We assume that the set of frequency channels is identical to the set of PUs. In the network routing problem, we assume that the set $\mathcal{N}$ is known, but we need to determine a set of $E$ links that optimizes the network utilities (to be discussed in Section III).
A SU node, $n_{i} \in \mathcal{N}$, establishes a link with its neighboring nodes.
Each PU $m_{k}\in \mathcal{M}$ is associated with a channel, which can be in  either an occupied  or
 an unoccupied state.
Let $\mathcal{S}_k, m_k\in\mathcal{M},$ be the set of channel states of PU $m_k$. A system state $s=[s_k]_{m_k\in\mathcal{M}} \in \mathcal{S}:=\prod_{k=1}^K\mathcal{S}_k$ is a collection of individual states of each primary channel $m_k$.

\begin{figure}\label{cr0}
    \centering
    \includegraphics[width=2.8in]{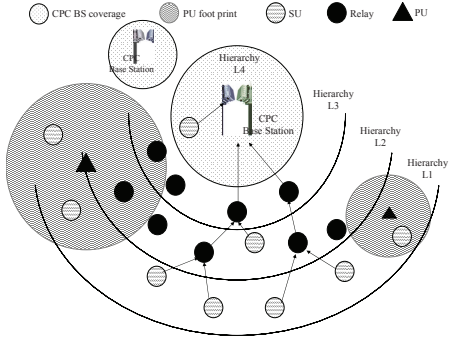}\vspace{-3.0mm}
    \caption{A snapshot of the proposed dynamic multi-hop network formation for the on-demand CPC on a frequency channel $m_{1}$.}
    \label{cr0}\vspace{-6.0mm}
\end{figure}

We consider that the CR network with SUs and relay nodes can be separated into a hierarchical structure  \cite{Han2009} using the medial axis \cite{Jiang2006} as a reference path as shown in Fig. \ref{cr0}.
Unlike \cite{Jiang2006}, particular physical channel conditions, such as delay constraints and spatial PUs' footprints, are required for multi-hop CR networks.
A SU has to connect to a CPC base station to request the CPC information through CR relay nodes in multi-hop CR networks.
Even though the SUs can be geographically distributed over the network, we can view the source nodes as nodes residing at the initial level $1$.
We let $\mathcal{L}(s)=\{1, 2, \cdots, L(s)\}$ be a set of hierarchical levels at state $s$, comprising a total of $L(s)$ levels with the first layer consisting of the source SUs and the terminal layer $L(s)$ consisting of nodes that directly connect to CPC base stations.
We define the medial axis between two PUs as the curve that describes the geographical points associated with SUs where the lowest power from PUs is perceived. We assume that each SU node is capable of sensing PUs and learning their footprints. The information of the footprint of PUs is supposed to be delivered to all SU nodes by flooding at the initialization of the SU network. SU nodes dynamically learn the pattern of the PUs' footprint. Denote by $A$ the medial axis and let $\mathcal{N}^{M}_{S_{A}}$ be the set of nodes on the medial axis $A$. We use the medial axis $A$ as the reference path along which the relay nodes between source nodes and CPC base stations are separated into $L(s)$ hierarchical levels. We suggest a relaxation factor, $\omega= {\vartheta}/{r}$, where $r$ is the radius of the canonical circle around the axis $A$ and the parameter $\vartheta$ is chosen between $0$ and  $1$. The relaxation factor allows us to increase the area around medial axis and consequently enlarge the area of SU nodes deployed in the network. Similarly, we use $\mathcal{N}^{r}_{S_{A}}$ to denote set of nodes in the relaxed area.

In an on-demand CPC scheme, each CPC base station can convey the statistics on the availability of idle frequency channels to the SU nodes to assist them in updating their spectrum knowledge. SU nodes access the nearest CPC base station to request the information. 
In our framework, we include the set of SUs $\mathcal{N}^{r}_{S_{A}}$ in the set $\mathcal{N}$, among which we determine the set of links $\mathcal{E}$ to connect SUs to CPCs. The remaining set $\mathcal{N}^{r}_{S_{A}}\backslash\mathcal{N}$ is the set of relay nodes along the medial axis that are capable of transmitting packets in a multi-hop fashion.

\section{Game Theory Model of Multi-hop Network}\label{sec:game_model}

In this section, we describe a stochastic multi-stage network routing game defined by $\boldsymbol{\Xi}=\{\Xi_h(s)\}_{s\in \mathcal{S}, h\in\mathcal{L}(s)}$, which is a set of games indexed by state  $s$ and hierarchy level $h$. Suppose the network maintains the same hierarchy structure at each state, then $\boldsymbol{\Xi}$ can be viewed as a matrix of games whose row indicates a spatial network routing game $\Xi(s)$ at a particular state $s$ and whose column is a temporal or state collection of games at the same level $h$ if at each state the network has the same hierarchical separation. The network formation game $\Xi(s)$ at a given state $s$ is well defined by a sequence of games $\{\Xi_{h}{(s)}\}_{h=1,\cdots, L(s)}$, where $\Xi_{h}{(s)}$ is a game at level $h$ and state $s$.
Each SU can access an idle PU's channel only when it is free. Let $\mathbb{P}_{k}:\mathcal{S}_{k}\times\mathcal{S}_{k}\rightarrow [0,1]$ be the state transition law of PU
channel $m_{k}$ on $\mathcal{S}_{k}$. Since the transition probabilities between states $x, y\in \mathcal{S}$ are only controlled by the PUs,  the stationary distribution $\pi=[\pi_{s}]_{s\in \mathcal{S}}$ of the Markov chain
$(\mathcal{S}_{k}, \mathbb{P}_{k})$, which we assume to exist, is independent of the actions of the SUs.
Let $\mathcal{H}_{l}, l\in\mathcal{L}(s),$ be the set of SUs that belong to the level $l$. It is clear that the sets $\mathcal{H}_l$ are mutually exclusive and $\cup_{l\in\mathcal{L}(s)}\mathcal{H}_l=\mathcal{N}$. 

Denote by $l_i\in\mathcal{L}(s)$ the level where user $n_i$ resides. To find a multi-hop connection to the CPC channel, user $n_i$ needs to find a node to connect to at the next level $l_i+1$. In this paper, the chosen node $(n_i, l_i+1)$ can potentially yield an optimal path with minimum payoff leading to a connection to the destination.  Let the set $\mathcal{N}_{l}(s)$ denote the nodes that are available for connection at the next level to a node at level $l$ at a particular state $s$. If  $l_i=L(s)$,  we let $\mathcal{N}_{l_i}=\mathcal{K}$.  Hence, $(n_i, l_i+1)$ is a node chosen by $n_i$ at the next level from the action set  $\mathcal{N}_{l_i}(s)$ available to node $(n_i,l_i)$. By default, we define $(n_i, l_i):=n_i$.

The local stage payoff to user $n_i$ is $u_{i}(s, (n_i, l_i+1), (n_{-i}, l_i+1)):\mathcal{S}\times \mathcal{N}_{l_i}^{|\mathcal{H}_{l_i}|}\rightarrow\mathbb{R}$. It is an instantaneous payoff at the local stage $l_i$ which depends on the local actions of all users at the level $l_i$. As a convention, we use $n_{-i}:=\mathcal{H}_{l_i}\backslash \{n_i\}$ to denote the set of users other than $n_i$ at the level $l_i$ and  $(n_{-i}, l_i+1):=\{(n_j, l_i+1): n_j\in \mathcal{H}_{l_i}\backslash\{n_i\}\}$ to mean the set of actions by the set of users $n_{-i}$. The coupling among the utility functions $u_{i}$ induces a noncooperative environment in which user $n_i$ competes with other user $n_{-i}$ to achieve optimal utilities.

In the multi-hop CR networks, spectrum occupancy is location-dependent, and therefore, along a multi-hop path, available spectrum channels may be different at each relay node \cite{Medley2009}. If the surrounding PUs are highly active, their availability for  communications duration becomes meager  for SU nodes, resulting in long routing delay in CR networks. Hence, we consider the queueing delay in the payoff function of the dynamic game. If a PU frequently intercepts the channels, the channels under current use need to be switched to other unoccupied channels and this switching time is added to the processing delay. In addition, if PUs intensively occupy all channels for a long time, SU nodes have no idle frequency channels and, consequently, re-routing delay is increased and results in more packet delay. The expected total packet delay $\tau_i$ perceived at SU node $n_{i}$ is defined by the Pollaczek-Khinchin formula for the M/G/1 queueing system as follows \cite{Chen1997}:
\begin{eqnarray}\label{TauCost}
\begin{aligned}
\lefteqn{\tau_i(s, (n_i,l_i), (n_{-i}, l_i))} \\
&= \frac{\lambda_{(n_i,l_i)}(s) \overline{X^{2}}_{(n_i, l_i)}(s)}{2(1-\rho(s, (n_i,l_i), (n_{-i}, l_i)) )}+\overline{X}_{(n_i, l_i)}(s),
\end{aligned}
\end{eqnarray}
where $\lambda_{(n_i,l_i)}(s)$ is the state-dependent arrival rate of packets from nodes at level $l_i$ seen at the node chosen by $n_i$ and $\rho(s, (n_i,l_i), (n_{-i}, l_i))  =\lambda_{(n_i,l_i)}(s)/\mu_{(n_i,l_i)}(s)=\lambda_{(n_i,l_i)}(s) \overline{X}_{(n_i, l_i)}(s)$, where $\mu_{(n_i,l_i)}(s)$ is the service time at the node $(n_i, l_i)$. $ \overline{X}_{(n_i, l_i)}$ is the mean service time per packet at the chosen node and $\overline{X^{2}}_{(n_i, l_i)}(s)$ is the expected variance of $ \overline{X}_{(n_i, l_i)}$. The coupling between transmitting nodes at level $l_i$ is evident from (\ref{TauCost}). When more nodes choose to connect to the same node $(n_i, l_i)$ as node $n_i$ does, they will experience more delay as the arrival rate $\lambda_{(n_i,l_i)}$ increases.

Each node $n_i$ calculates its payoff $u_i(s, (n_i,l_i), (n_{-i}, l_i))$ of connecting to $(n_i, l_i)$ given by
\begin{equation}\label{t_tutil}
u_i(s, (n_i,l_i), (n_{-i}, l_i)) =\tau_i(s, (n_i,l_i), (n_{-i}, l_i)).
\end{equation}
The degree of freedom of user $n_i$'s choice is constrained by the set $\mathcal{N}_{l_i}$.
Given a state $s$, a user aims to optimize his long-term payoff along his path to a CPC base station  rather than merely to optimize his local myopic payoff $u_i$ at level $l_i$. Denote by  $\mathcal{P}(n_i, l), l_i<l\leq L(s)$,  the path of a node $n_i$ to a node at level $l$. It is clear that when $l=L(s)$, $\mathcal{P}(n_i, l)$ refers to the path from a SU to the destination and, by default, when $l=l_i+1$, $\mathcal{P}(n_i, l_i+1)$ is a link from $n_i$ to the node $(n_i, l_i+1)$. For example, the path $\mathcal{P}(n_i, l_i+2)$ is composed of two links: $\mathcal{P}(n_i, l_i+1)$ and $\mathcal{P}((n_i,l_i+1), l_i+2)$. The user $n_i$ can only have control over the link $\mathcal{P}(n_i, l_i+1)$ while the link $\mathcal{P}((n_i,l_i+1), l_i+2)$ and the links onwards are controlled by other nodes. 

Let $U_i(s, (n_i, l), (n_{-i}, l)): \mathcal{S}\times \prod_{l'\in\{l_i, l_i+1, \cdots, l\}}\mathcal{N}_{l'}^{|\mathcal{H}_{l'}|}\rightarrow\mathbb{R}$ denote the path utility attained by the user $n_i$ at state $s$ over the path $\mathcal{P}(n_i, l)$. Due to the delay, interference and coupling, the utility also depends on the path by other users. Following the convention, we use $(n_i,l):=\{((n_i,l'),l'+1): l_i\leq l'\leq l-1\}$ to denote the path of user $n_i$ up to the level $l$ and use
$(n_{-i}, l):=\{((n_j,l'), l'+1):$ $n_j\in \mathcal{H}_{l_i}\backslash\{n_i\}, l_i\leq l' \leq l-1 \}$ to denote the actions by other users up to the level $l$ along their chosen path $\mathcal{P}(n_{-i}, l)$.
The goal of a SU $n_i$ is to attain a connection to a CPC base station with optimal utility along the path.
The path utility  $U_i(s, (n_i, L(s)), (n_{-i}, L(s)))$ of a user $n_i$  over the path $\mathcal{P}(n_i, L(s))$ can be expressed as a sum of stage utilities as follows:
\begin{eqnarray}\label{U1}
\begin{aligned}
\lefteqn{U_{i}(s,(n_i, L(s)),(n_{-i}, L(s)))}\\
&=\sum_{l=l_i}^{L(s)} u_{(n_i, l)}(s, (n_i, l+1), (n_{-i}, l+1)),
\end{aligned}
\end{eqnarray}
where and $u_{(n_i, l)}(s, (n_i, l+1),  (n_{-i}, l+1))$ denotes the payoff to node $(n_i,l)$ when it chooses directly the node $(n_i, l+1)$ in the next level.

At a fixed state $s$, a SU  $n_i$ needs to choose an optimal path that achieves the optimal payoff. However, he can only choose from $\mathcal{N}_{l_i}(s)$ his next hop connection $(n_i,l_i+1)$ and leaves the future choices made by the node that he connects to. Let $U^*_i(s)$ be the optimal payoff. The path utility (\ref{U1}) can be rewritten as 
\begin{eqnarray}\label{U22}
\begin{aligned}
\lefteqn{U^*_i (s, (n_{-i}, L(s))} \\
&:= \min_{(n_i, l_i+1)\in\mathcal{N}_{l_i}(s)} \sum_{l=l_i}^{L(s)} u_{(n_i, l)}(s, (n_i, l+1), (n_{-i}, l+1))\\
&= \min_{(n_i, l_i+1)\in\mathcal{N}_{l_i}(s)} u_i(s, (n_i, l_i+1), (n_{-i}, l_i+1))\\
&\;\;\;\;+ \sum_{l=l_i+1}^{L(s)}u_{(n_i, l)}(s, (n_i, l+1), (n_{-i}, l+1))\\
&= \min_{(n_i, l_i+1)\in\mathcal{N}_{l_i}(s)} u_i(s, (n_i, l_i+1), (n_{-i}, l_i+1))\\
&\;\;\;\;+ U^*_{(n_i, l+1)}\left(s,(n_{-(n_i, l+1)}, L(s))\right) \\
&= \min_{(n_i, l_i+1)} u_i(s, (n_i, l+1))+ U^*_{(n_i, l+1)}(s,(n_{(n_i, l+1)}, L(s))).
\end{aligned}
\end{eqnarray}

On the right-hand side of (\ref{U22}), $U^*_{(n_i, l+1)}$ can be seen as the payoff-to-go and $u_i$ is the current instantaneous payoff to optimize. A solution to (\ref{U22}) can be found using backward induction, where we start with the nodes at the last level and then propagate the solution to level $l_i$. Since every user on the same level with $n_i$ optimizes his utility in the same way, the best responses of (\ref{U1})  to other nodes on the same level can lead to a Nash equilibrium where no node finds it to its benefit to deviate from its chosen action. However, the existence of Nash equilibrium  for the game at a particular level is not guaranteed. To ensure the existence, we  adopt mixed strategies in which the users at level $l_i$ randomize over $\mathcal{N}_{l_i}(s)$.  Let $\mathbf{f}_{i, l}(s)=\{\mathbf{f}_{(n_i,l')}: l_i\leq l'\leq l\}$ denote a mixed strategy of $n_i$ at a given state $s$ up to level $l$. It is clear that when $l=l_i$, then $\mathbf{f}_{i, l}(s)$ only contains the mixed strategies of node $n_i$. For $l> l_i$, $\mathbf{f}_{i, l}(s)$ comprises of sets of mixed strategies along the path up to $(n_i, l)$.  We let $\mathbf{F}_l:= \{\mathbf{f}_{i,l'}(s), n_i\in\mathcal{H}_{l_i}, l_i \leq l'\leq l\}$  the set of user strategies at level $l$ and $\mathbf{F}_{-i,l}:= \{\mathbf{f}_{j,l'}(s), n_j\in\mathcal{H}_{l_i}\backslash\{n_i\}, l_i \leq l'\leq l\}$. Let $\mathbf{F}:= \{\mathbf{f}_{i,l}(s), n_i\in\mathcal{H}_l, l\in\mathcal{L}(s)\}$.

A user $n_i$ chooses a mixed strategy to minimize his expected total payoff of (\ref{U1}), i.e.,
\begin{equation}\label{EU1}
\overline{U}_i(s, \mathbf{F})=\sum_{l=l_i}^{L(s)} \mathbb{E}_{\mathbf{f}_{i,l}, \mathbf{F}_{-i,l}} u_{(n_i, l)}(s, (n_i, l+1),  (n_{-i}, l+1)).
\end{equation}
Following (\ref{eqn1}), we have
\begin{eqnarray}\label{eqn1}
\begin{aligned}
\lefteqn{\overline{U}^*_i\left(s, \mathbf{f}_{i,l_i} \mathbf{F}_{-i,l_i}\right)} \\
&:= \min_{\mathbf{f}_{i, l_i}} \mathbb{E}_{\mathbf{f}_{i, l_i}, \mathbf{F}_{-i, l_i} } u_i(s, (n_i, l_i+1),(n_{-i}, l_i+1)) \\
&\;\;\;\; +\overline{U}^*_{(n_i, l+1)}\left(s,\mathbf{F}_{(n_i, l+1), l+1}\right).
\end{aligned}
\end{eqnarray}
Similarly, the game with expected payoff can also be solved by backward induction. At each state,  a matrix game needs to be solved and each user $n_i$ generates a mixed strategy on their action set.
Hence, what we will be encountering here is a Nash equilibrium in {\em behavioral} strategies.
\section{Dynamic Multi-hop Routing}\label{sec:dynamic model}
A user $n_i$  faces a long-term payoff when the cognitive radio system evolves on the state space $S$. Let $\{s_t, t\in \mathbb{Z}\}$ be a sequence of states indexed by time $t$. Let $v_i(s)$ be the value function of user $n_i$ when $n_i$ starts in state $s$, i.e., $s_0=s$.  We consider mixed (behavioral) stationary strategies for user $n_i$ that are only dependent on their current state. Denote by $f_i(s, a_i)$ the probability of user $n_i$ choosing a next level node $a_i\in\mathcal{N}_{l_i}$ at state $s\in\mathcal{S}$. The vector $\mathbf{f}_i(s)\in[0,1]^{|\mathcal{N}_{l_i}|}$ is the state dependent mixed strategy given by $\mathbf{f}_i(s)=[f_i(s, a_i)]_{a_i\in\mathcal{N}_{l_i}}$. Let $\mathcal{F}_i(s)$ denote the set of all such strategies, i.e., 
\begin{eqnarray}
\mathcal{F}_i(s)=\left\{\mathbf{f}_i(s)\in[0,1]^{|\mathcal{N}_{l_i}|}~\vline~ \sum_{a_i\in\mathcal{N}_{l_i}}f_i(s, a_i)=1\right\}.
\end{eqnarray}
Let $V_i$ be the long-term infinite horizon path utility function which depends on  $\mathbf{f}_i(s), n_i \in \mathcal{H}_{l_i}$; it is given by 
{\small
\begin{flushleft}
$V_i(s, \mathbf{f}_i(s), \mathbf{f}_{-i}(s))$
\end{flushleft}
\begin{align}\label{Vi}
=\sum_{t=0}^\infty \beta^t \mathbb{E}_{s,\mathbf{f}_i(s), \mathbf{f}_{-i}(s)}[U_i(s, (n_i, L(s)),(n_{-i}, L(s))) | s_0=s] \ \ \ \ \ \ \ \ \ \ \ \ \\
\nonumber =\sum_{t=0}^\infty\sum_{l=l_i}^{L(s)}  \beta^t \mathbb{E}_{s,\mathbf{f}_i(s),\mathbf{f}_{-i}(s)}[u_{(a_i, l)}(s, (n_i, l+1), (n_{-i}, l+1)) | s_0=s], \
\end{align}}
where $0 < \beta < 1$ is a discounting factor. The value function is yielded by the optimization over the long term utility $V_i$ given the other users' mixed stationary strategies $\mathbf{f}_{-i}^*(s)$,
\begin{eqnarray}
v_i(s)=\min_{\mathbf{f}_i\in\mathcal{F}(s)} V_i(s, \mathbf{f}_i(s), \mathbf{f}_{-i}^*(s)), \forall n_i\in\mathcal{H}_{l_i}.
\end{eqnarray} 

Since the utilities $u_i$ are bounded, $V_i$ is bounded and the interchange of expectation and summation is possible by Fubini's Theorem \cite{basar95}. Assume $L(s)=L$ for all $s\in\mathcal{S}$. We can rewrite (\ref{Vi}) as 
{\small
\begin{eqnarray}
\nonumber\begin{aligned}
\lefteqn{V_i(s, \mathbf{f}_{i,l_i}(s), \mathbf{F}_{-i,l_i}(s))} \\
&=\sum_{l=l_i}^{L}\sum_{t=0}^\infty \beta^t \mathbb{E}_{s,\mathbf{f}_{i,l}(s), \mathbf{F}_{-i,l}(s)} [u_{(n_i, l)}(s, (n_i, l+1), (n_{-i}, l+1))| s_0=s] \\
&=\sum_{t=0}^\infty  \beta^t \mathbb{E}_{s,\mathbf{f}_{i,l_i}(s), \mathbf{F}_{-i,l_i}(s)}[u_{i}(s, (n_i, l_i+1), (n_{-i}, l_i+1)) |s_0=s] \\
&+\sum_{l=l_i+1}^{L}\sum_{t=0}^\infty \beta^t \mathbb{E}_{s,\mathbf{f}_{i,l}(s), \mathbf{F}_{-i,l}(s)}[u_{(n_i, l)}(s, (n_i, l+1),(n_{-i}, l+1))|s_0=s].
\end{aligned}
 \end{eqnarray}}
Denote by $W_{i} (s, \mathbf{f}_{i} (s),\mathbf{F}_{-i,l_i} (s))$ the local infinite-horizon utility function, i.e., 
\begin{eqnarray}
\begin{aligned}
\lefteqn{W_{i} (s, \mathbf{f}_{i,l_i} (s)), \mathbf{F}_{-i,l_i} (s))}\\
&= \sum_{t=0}^{\infty} \beta^{t} \mathbb{E}_{s,\mathbf{f}_{i,l_i}(s), \mathbf{F}_{-i,l_i}(s)} [u_{i}(s, (n_{i}, l_{i+1}), (n_{-i}, l_{i+1}))|s_{0} = s].
\end{aligned}
\end{eqnarray}
Hence, (\ref{Vi}) can be rewritten as
\begin{eqnarray}
\begin{aligned}
\lefteqn{V_i(s, \mathbf{f}_{i,l_i}(s), \mathbf{F}_{-i,l_i}(s)) = W_i(s, \mathbf{f}_{i,l_i}(s),  \mathbf{F}_{-i,l_i}(s))} \\
&+\sum_{l=l_i+1}^{L} W_{(n_i, l)} (s, \mathbf{f}_{(n_i, l)}(s),  \mathbf{F}_{(-(n_i, l), l)}(s)).
\end{aligned}
\end{eqnarray}
The payoff function $V_i$ has two components. One is the local infinite time horizon payoff and the other is the off-to-go infinite time horizon payoff. Both components depend on the strategy made by user $n_i$.

 \subsection{Nash Equilibrium and Backward Induction}
We intend to find the Nash equilibrium of the game defined by $\boldsymbol{\Xi}$ with the utility function in (\ref{Vi}).  
 \begin{definition}
Let $\Phi_{l}$ be the level $l$ game denoted by $\Phi_l:=\langle \mathcal{N}_l, \{V_i, n_i\in\mathcal{N}_l\}, \{\mathcal{A}_i, n_i\in\mathcal{N}_l\}, \mathbb{P}, \mathcal{S}\rangle$, where $V_i$ is the utility defined in (\ref{Vi}) and $\mathcal{A}_i=\mathcal{A}_l=\mathcal{N}_{l}=\mathcal{H}_{l+1}, \forall n_i\in\mathcal{H}_l$ is the action space for users $n_i\in\mathcal{N}_l$. The mixed stationary strategy $\mathbf{F}^*$ is a Nash equilibrium if it satisfies $\forall n_i\in\mathcal{N}_{l_i},$
$$
{V_i(s, \mathbf{F}^*_{l_i}(s))}
\geq V_i(s, \mathbf{f}_i(s), \mathbf{F}^*_{-i, l_i}(s)), \forall \mathbf{f}_{i,l_i}(s)\in\mathcal{F}_i(s).
$$
The sequence of $\{\Phi_l\}_{l\in\mathcal{L}}$ admits a mixed stationary Nash strategy $\mathbf{F}^*$ if for all $l\in\mathcal{L}$, $\mathbf{F}^*_l$ is a mixed stationary Nash strategy for the game $\Phi_l$.
 \end{definition}

The value function $v_i$ is the payoff at the Nash equilibrium, i.e.,
 $ v_i(s)=  V_i(s, \mathbf{F}^*_{l_i}(s)).$ In short-hand notation, we use 
$
 \mathbf{f}_i^* \in \textrm{arg \ NE}_i \{V_i(s, \mathbf{F}_{l_i}(s))\}, \
 v_i=\textrm{NE}_i  \{V_i(s, \mathbf{F}_{l_i}(s))\},
$
to denote the Nash equilibrium and its corresponding value function respectively.
Using induction, we have the following result.
\begin{theorem}
Let $r_{i,l_i}(s)=\textrm{NE}_i \{u_{i}(s, (n_i, l_i+1),(n_{-i}, l_i+1))\}, n_i\in\mathcal{H}_{l_i}$ and $\mathbf{r}_{i,L}$ be its vector form with each entry corresponding to one state. Also let $\mathbf{P}=[P_{ss'}]_{s, s'\in \mathcal{S}}$ be the transition matrix.
At the last stage, $l_i=L$, we obtain
 \begin{eqnarray}
 \nonumber \mathbf{f}_{i,l_i}(s) &\in&\textrm{arg \ NE}_i \{ u_{i}(s, (n_i, l_i+1),(n_{-i}, l_i+1)) \}\\
\nonumber \mathbf{v}_i&=&[\mathbf{I}-\beta\mathbf{P}]^{-1}\mathbf{r}_{i,L},
 \end{eqnarray}
 where ${r}_{i,L}(s)=\textrm{NE}_i \{u_{i}(s, (n_i, L),(n_{-i}, L))\}$. If $\mathcal{K}=\{K_0\}$ is a singleton set, then  ${r}_{i,L}(s)= u_{i}(s, K_0,\cdots, K_0)$.
The mixed stationary strategies $\mathbf{F}^*$ of the game $\boldsymbol{\Xi}$ can be found recursively by
\begin{eqnarray}
\nonumber \mathbf{f}_{i,l_i}^*(s)&\in & \textrm{arg \ NE}_i\{u_i(s, (n_i, l_i+1), (n_{-i}, l_i+1))\\
\nonumber &&+ v_{(n_i, l_i+1)}(s)\}, n_i\in\mathcal{H}_{l_i}.
\end{eqnarray}
The optimal payoff to a node $n_i$ playing $\mathbf{f}^*_{i,l_i}(s)$ yields a value function
\begin{eqnarray}
 \mathbf{v}_i =[\mathbf{I}-\beta\mathbf{P}]^{-1}
 \mathbf{r}_{i,l_i}
, \forall n_i\in\mathcal{H}_{l_i},
\end{eqnarray}
where
\begin{eqnarray}
\nonumber \begin{aligned}
r_{i, l_i}(s)=
\left\{
\begin{array}{ll}
 \textrm{NE}_i\{u_i(s, (n_i, l_i+1), (n_{-i}, l_i+1))
 \\ \ \ \ \ \  +v_{(n_i, l_i+1)}(s)\}  \ \ \ \ \ \ \ \ \ \ \textrm{~if~}  1 \leq l_i\leq L-1,
\\  \textrm{NE}_i\{u_i(s, (n_i, l_i+1), (n_{-i}, l_i+1))\} \textrm{~if~}  l_i=L.
\end{array}
\right.
\end{aligned}
\end{eqnarray}
 \end{theorem}
The recursion involves finding $\mathbf{r}_{i,l_i}$ and $\mathbf{f}^*_i(s)$ through the value functions of nodes at the next level.

\subsection{Algorithm Description}
\begin{table}
  \centering
  \caption{The skeleton of the proposed dynamic routing algorithm}\label{t:algorithm}
  \begin{tabular}{l}
    \hline
 \ \ \ \ \textbf{1: Global procedure} \\
\ \ \ \ \ \ \ \ Initialize network \\
\ \ \ \ \ \ \ \ \ Define $\partial R$, $A$, $\omega$,  $\mathcal{N}^{r}_{SA}$  \\
\ \ \ \ \ \ \ \ Update map \\
\ \ \ \ \textbf{2: Local procedure} \\
\ \ \ \ \ \ \ Start game $\Xi(s)$ \\
\ \ \ \ \ \ \ \ \ SUs learn the mixed strategies by fictitious play at each level \\
\ \ \ \ \ \ \ End game $\Xi(s)$ \\
\ \ \ \  \textbf{3: Packet transmission procedure} \\
\ \ \ \ \textbf{4: Complete forwarding the request of CPC information} \\
\ \ \ \ \textbf{5: Acquire information from CPC} \\
\ \ \ \ \textbf{6: Go to update map if new information is obtained} \\

\hline
  \end{tabular}
\end{table}\vspace{-0.5mm}
We propose a distributed dynamic routing algorithm for the SUs to minimize their interference with the PUs. Each node improves its current payoffs and takes into account the previous payoffs. Summarized in Table I,  the algorithm starts with a global procedure defining the set of hierarchies from the sources to the destinations. In this step, we initialize the boundary, PU footprint map, and the medial axis and they are dynamically updated whenever the SUs acquire the new system knowledge from CPCs. Once the source to the destination pair is defined, the set of $\mathcal{N}^{r}_{S_{A}}$ along the medial axis is defined as a reference routing path with the relaxation factor $\omega$. The second step of the algorithm is a local procedure in which each user calculates his payoff and selects the best routing node at the next level with the minimum delay.
The node updates its mixed strategies by  fictitious play \cite{Fudenberg98} until the game converges to its Nash equilibrium. We use  fictitious play at each stage $l\in \mathcal{L}$ to find the mixed Nash equilibrium at that level. After SUs acquire the state information from CPCs, the nodes update map including the PUs' map and topology information and the set of nodes $\mathcal{N}^{M}_{SA}$ and $\mathcal{N}^{r}_{SA}$.


\section{Simulation Results}\label{sec:sim_results}
\begin{figure*}[t]
\begin{center}
\begin{minipage}[b]{0.30\linewidth}\label{fig:ne1}
\centerline{\psfig{figure=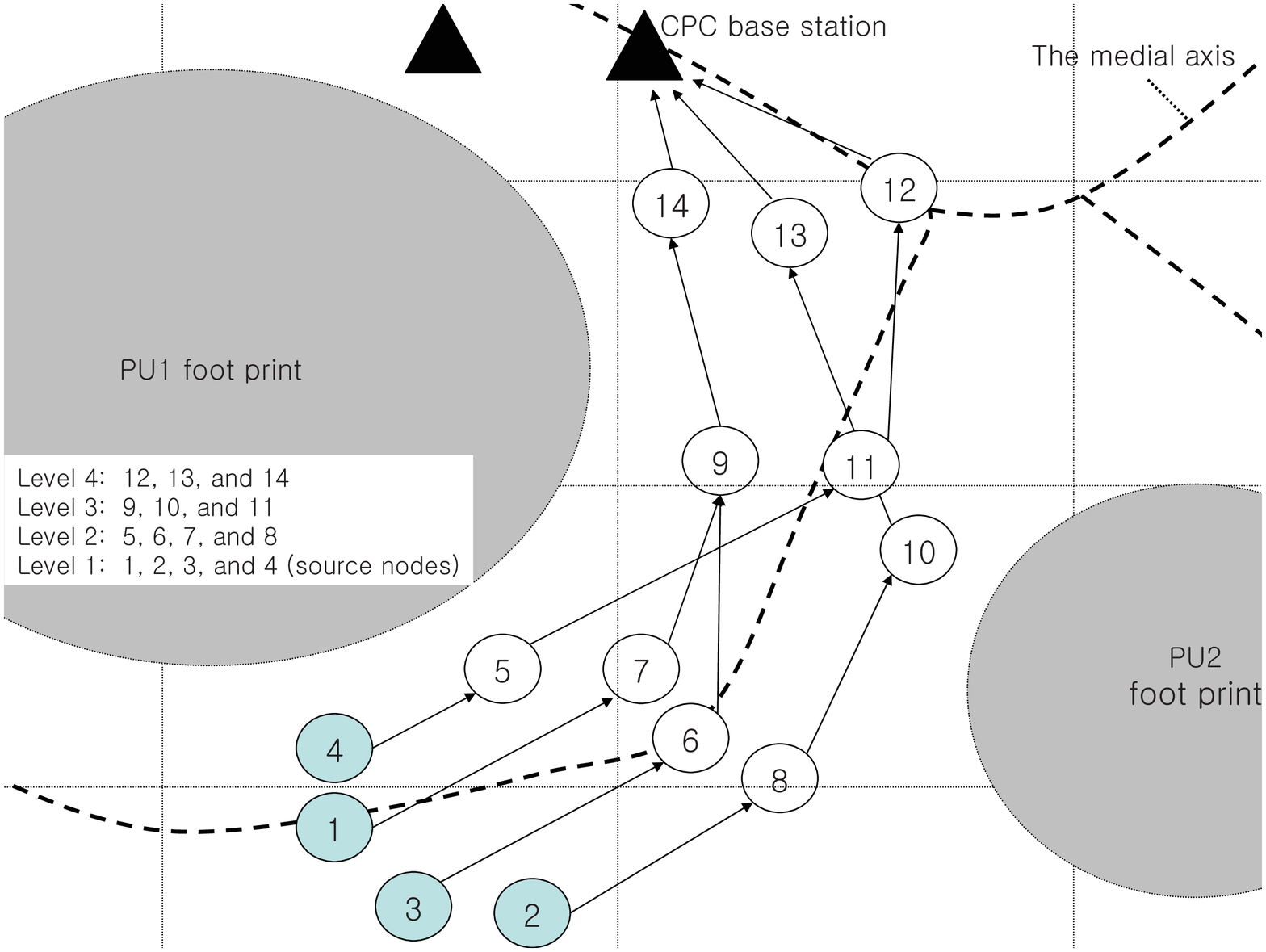, scale=0.228}} \caption{Results of
the  dynamic routing algorithm with 4 SU nodes and 10 relays
randomly deployed in $\mathcal{N}^{r}_{SA}$ at $t=t^{'}$. $\omega =
0.7$ and random $U$ applied for every nodes. The arrowed solid lines
are final routing results, i.e. Nash equilibrium for each SU nodes.}
\label{fig:ne1}
\end{minipage}
\hspace{0.3cm}
\begin{minipage}[b]{0.30\linewidth}\label{fig:dnodes} 
\centerline{\psfig{figure=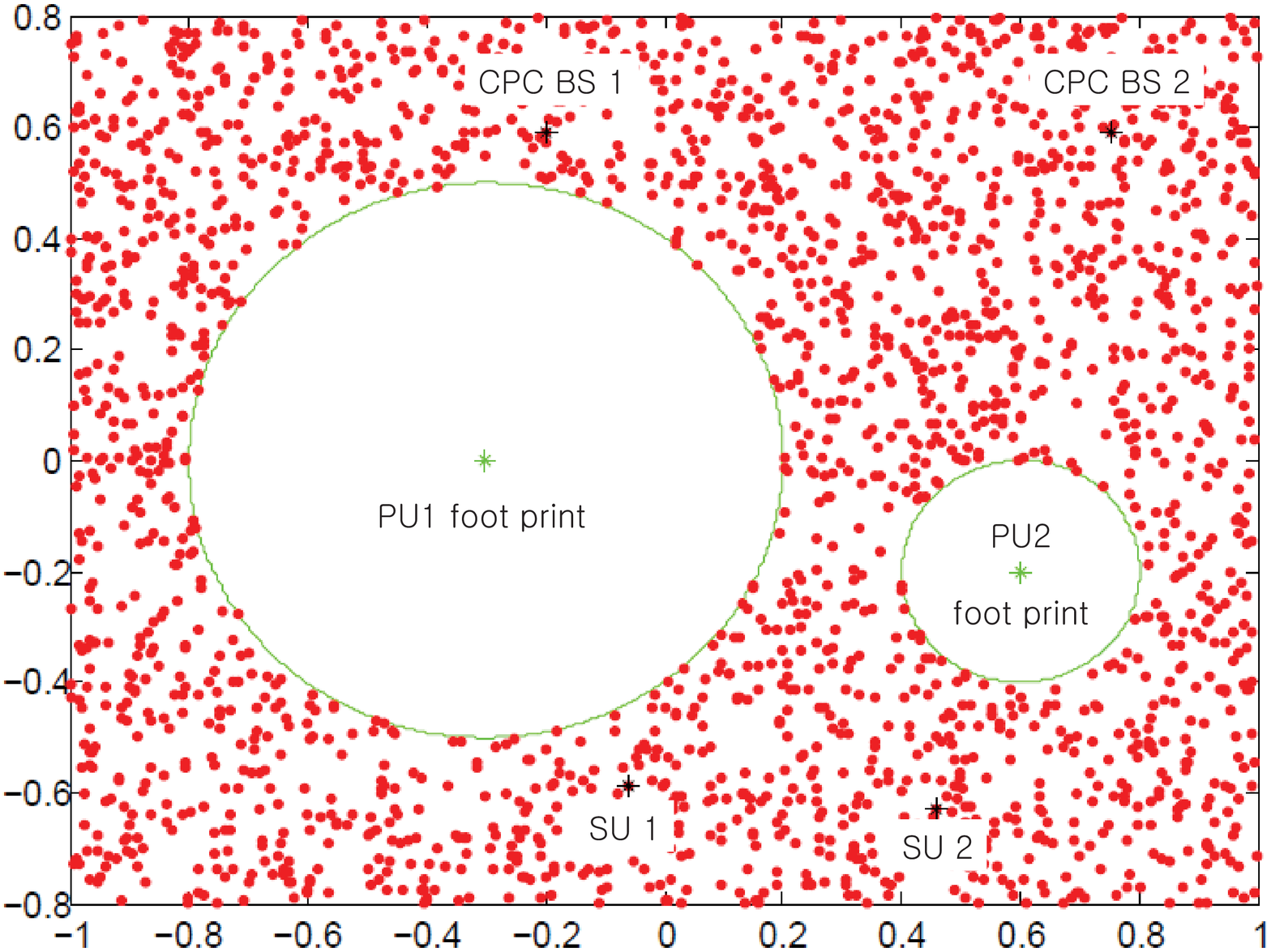, scale=0.228}} \caption{The
network topology for 3,000 SU nodes and relays (dot) randomly
deployed in the presence of 2 PUs.}\label{fig:dnodes}
\end{minipage}
\hspace{0.3cm}
\begin{minipage}[b]{0.30\linewidth}\label{fig:droutes}
\centerline{\psfig{figure=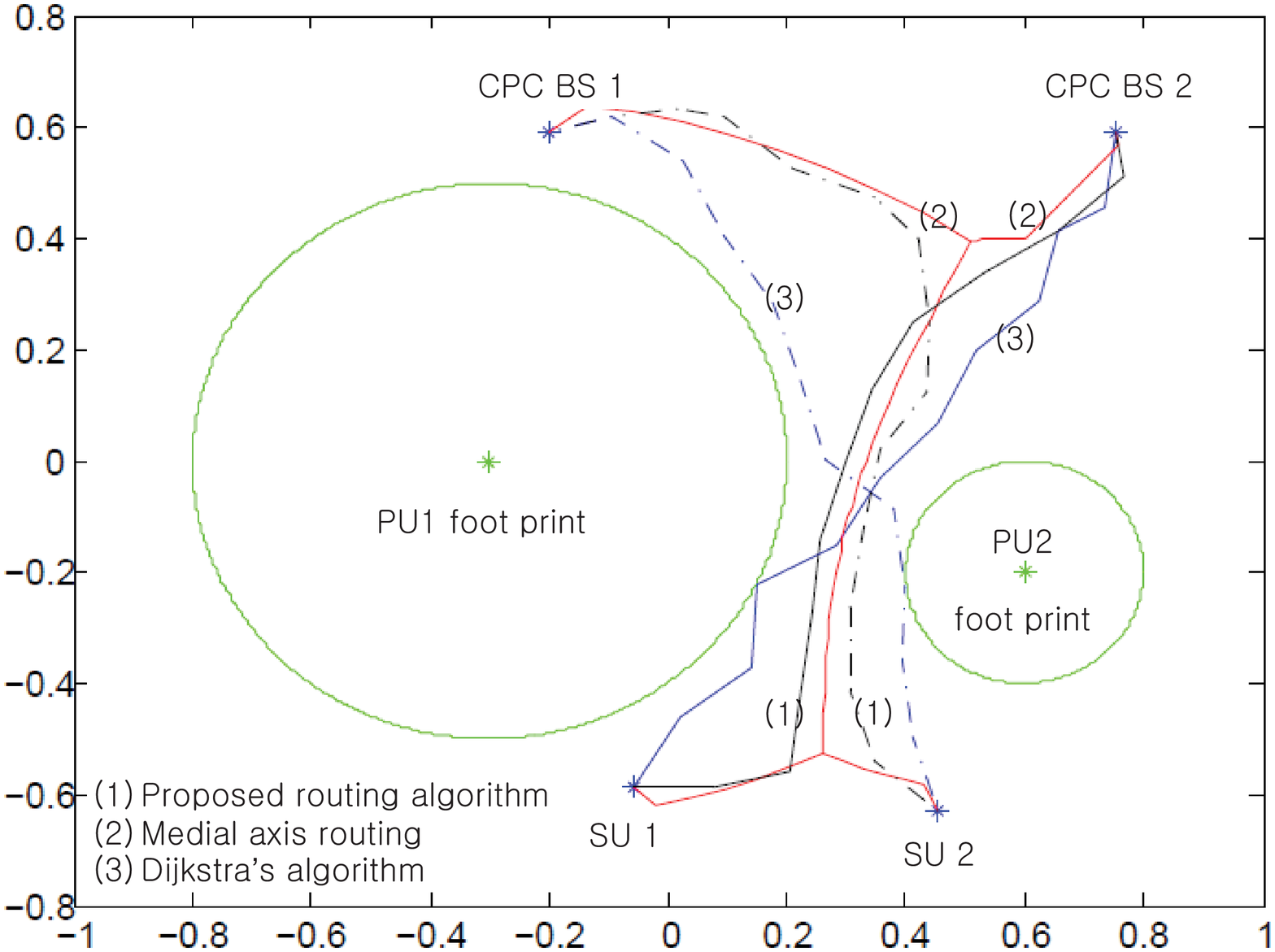, scale=0.228}}
\caption{Results of routing for the proposed algorithm in comparison
with Dijkstra's algorithm and MA routing. $\omega = 0.7$ and random
$U$ deployed for every SU nodes.\vspace{-5mm}} \label{fig:droutes}
\end{minipage}
\end{center}\vspace{-5mm}
\end{figure*}
\begin{figure*}[t]
\begin{center}
\begin{minipage}[b]{0.30\linewidth}\label{fig:converge1}
\centerline{\psfig{figure=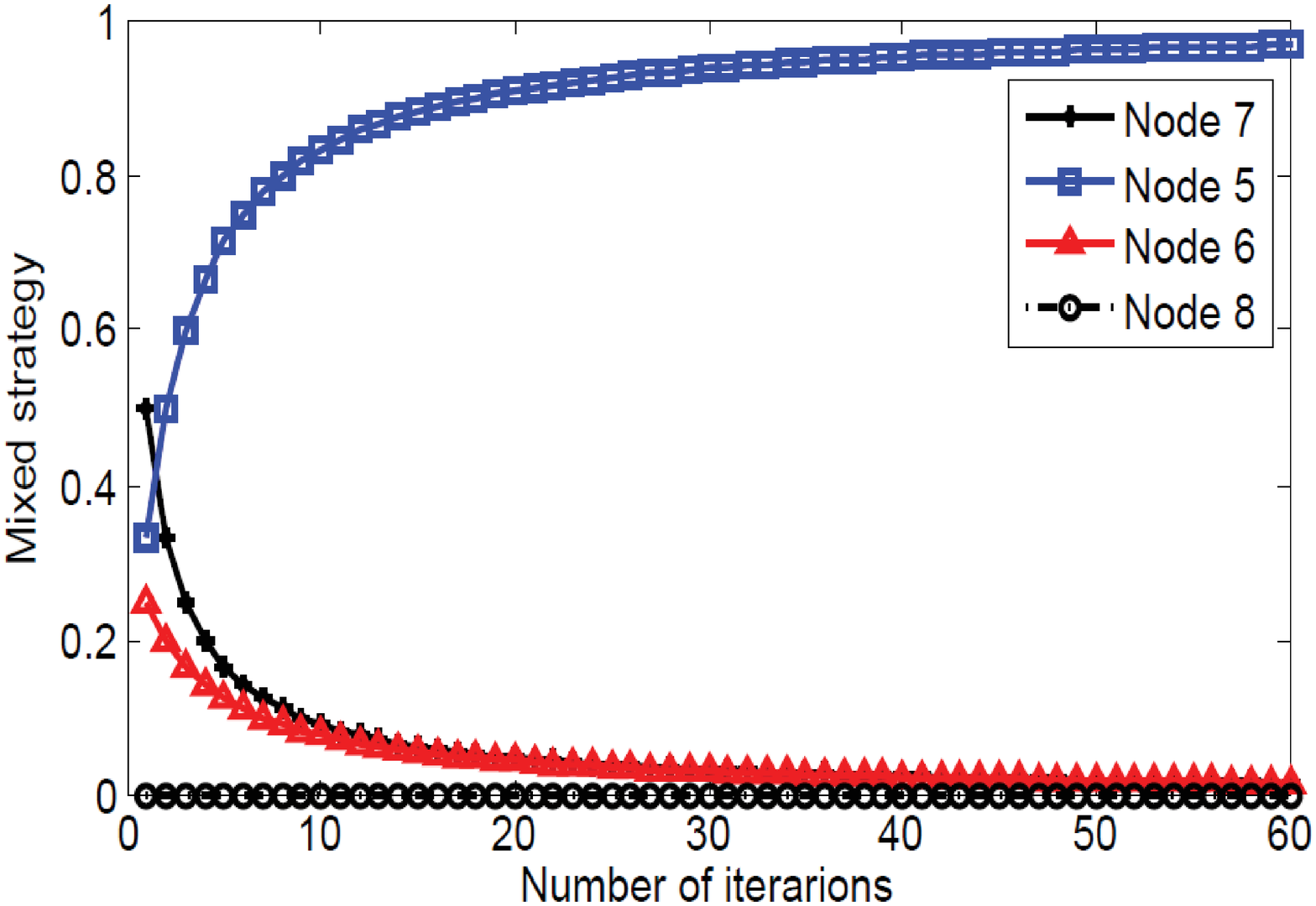, scale=0.228}}\vspace{-3mm}
\caption{Convergence of the mixed strategies vs. iteration time of
node 4 at the first hierarchy level in Fig. \ref{fig:ne1}.}
\label{fig:converge1}
\end{minipage}
\hspace{0.3cm}
\begin{minipage}[b]{0.30\linewidth}\label{fig:int2} 
\centerline{\psfig{figure=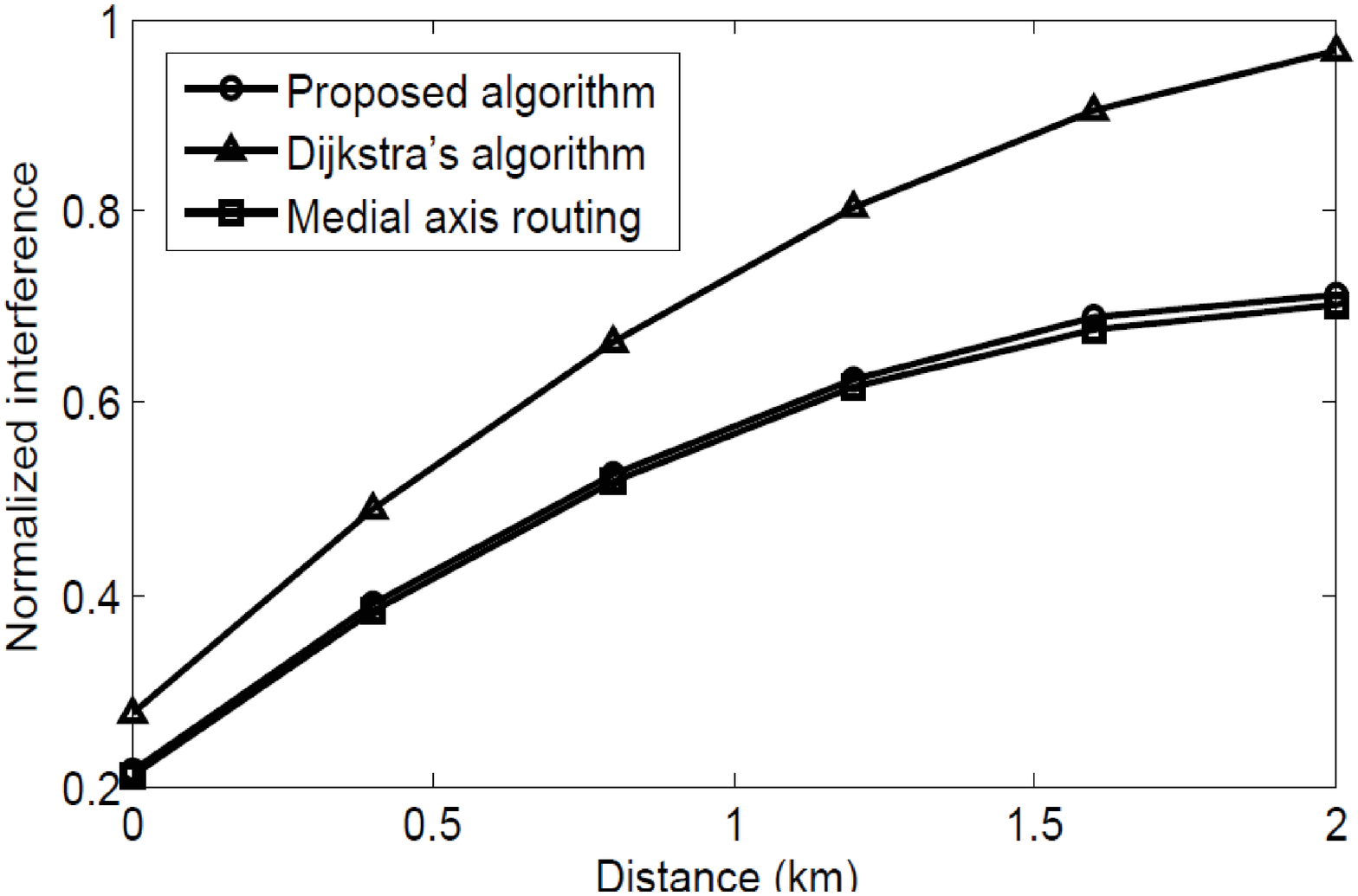, scale=0.228}}\vspace{-3mm}
\caption{The normalized interference results versus distance of
routing paths for different algorithms.}\label{fig:int2}
\end{minipage}
\hspace{0.3cm}
\begin{minipage}[b]{0.30\linewidth}\label{fig:delay2}
\centerline{\psfig{figure=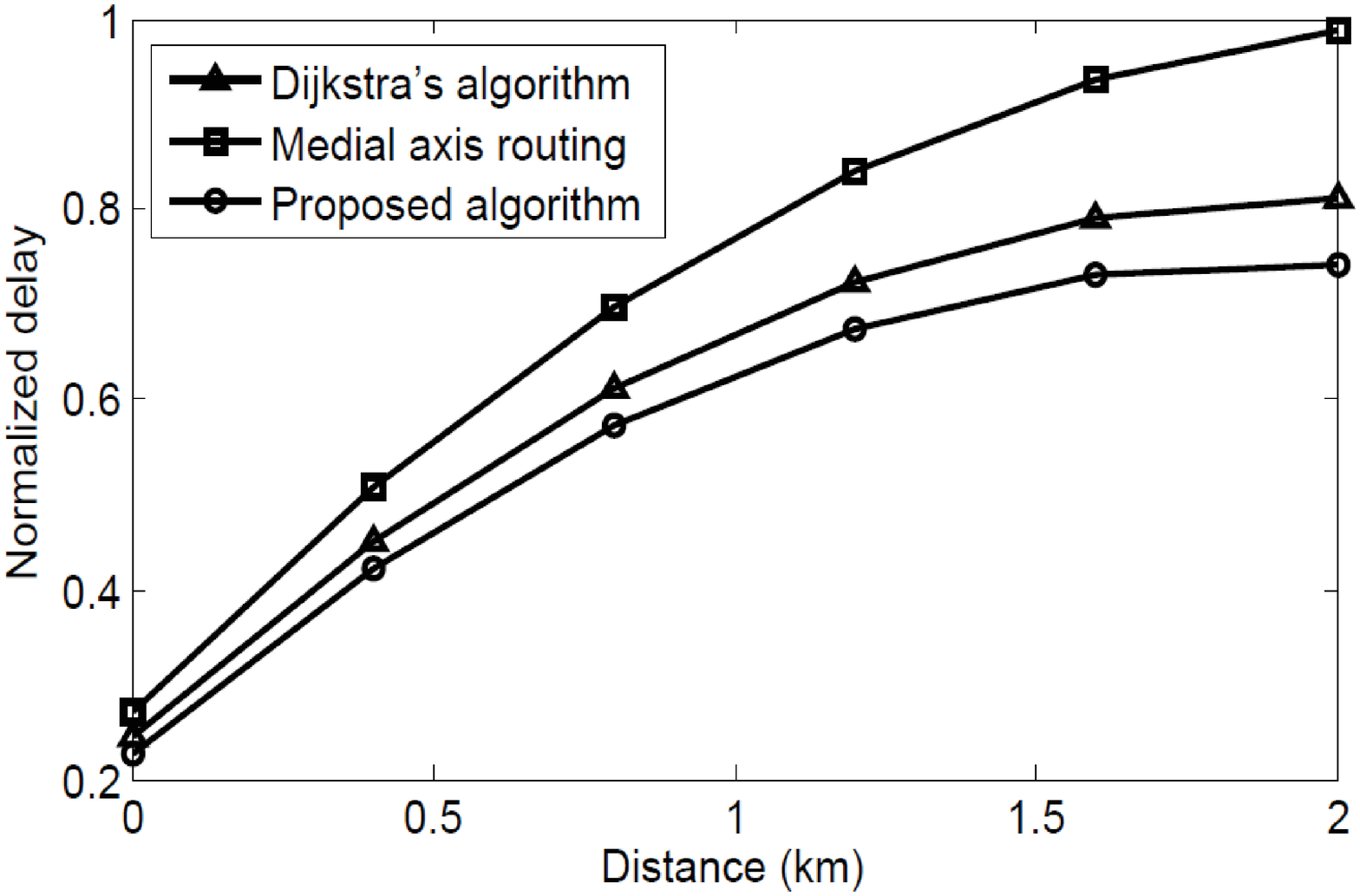, scale=0.228}}\vspace{-3mm}
\caption{The normalized delay results versus distance of routing
paths for comparing with different routing
algorithms.\vspace{-10mm}} \label{fig:delay2}
\end{minipage}
\end{center}\vspace{-5mm}
\end{figure*}

Fig. \ref{fig:ne1} shows the simulation results of a routing
formation from 4 source SU nodes to the nearest CPC base station within
$\mathcal{N}^{r}_{SA}$. In Fig. \ref{fig:ne1}, we show
two PUs, two CPC base stations, 4 source SUs, and 10
SU relay nodes are in $\mathcal{N}^{r}_{SA}$ within a 2 km $\times$
2 km area.  We assume that the PUs do not vary their configurations or footprints.
The SU source nodes are 1, 2, 3, and 4 at the first
hierarchy level. SU relay nodes have their own random arrival time
and service time. The arrival time depends on routing results at
previous hierarchy levels due to the dynamic utility functions. The
nodes update their mixed strategies at each iteration. The optimal
action taken at the iteration $k$ determines the vector $U_{i}$ at the
following iteration and is used to update the empirical frequencies. The SUs form the current belief of a user on other users' actions from the empirical learning and select the best response action in the next iteration. This iterative process continues until the empirical
frequencies converge to the Nash equilibrium.

Fig.
\ref{fig:ne1} depicts 4 SU nodes, i.e., 1, 2, 3, and 4, at the first level and 4
SU relay nodes, i.e., 5, 6, 7, and 8, at the next level. Fig. \ref{fig:converge1} shows the convergence of the mixed strategies vs. iterations at SU node 4 at the first
hierarchy level. The node 4 updates its mixed strategies by
increasing or decreasing the probabilities on selecting nodes  5,
6, 7, and 8, as the data of the play history accumulates. The mixed
strategies finally converge to 0.9, 0.2, 0.01, and 0.01 for
connecting to nodes 5, 6, 7, and 8 within 60 iterations. Therefore,
node 4 selects node 7 to transmit packets within a reasonable
iteration time.

The results of the dynamic routing algorithm for densely
deployed SU networks are shown in Fig.\ref{fig:droutes}. There are
3,000 nodes and relays randomly deployed within a 1.6 km by 2.0 km area in Fig.
\ref{fig:dnodes}. Two PUs exist respectively at (-0.3, 0) with the
coverage radius of 0.5 km and at (0.6. -0.2) with the coverage
radius of 0.2 km. The arrival rate and the processing rate for all
SU nodes are randomly chosen. Fig.
\ref{fig:droutes} shows two pairs of routing results from SU 2 to CPC
BS 1 and SU1 to CPC BS 2. In this simulation, all SU nodes have 1 W
transmission power with the interference range of 0.15 km and a path
loss factor $\alpha=2.5$. The results of the proposed routing
algorithm with $\omega=0.7$ is compared with results of Dijkstra's
algorithm and medial axis (MA) routing as shown in Fig.
\ref{fig:droutes}. The blue lines (3) are routes using Dijkstra's
algorithm, and the black lines (1) represent the routes using our
algorithm. The red lines (2) show the routes using MA algorithm.
Intuitively, the Dijkstra's algorithm provides higher interference
with PU networks than the proposed algorithm along overall routing
paths and MA routing algorithm provides much higher delay because of
traffic jam on medial axis path.

Fig. \ref{fig:int2} shows the results of interference comparison
between different algorithms. The x-axis represents the distance
between the source and destination and y-axis is the normalized
interference. The interference with PUs is calculated using the
interference temperature model. The average delay results versus
distance of routing paths is also shown in Fig. \ref{fig:delay2}.
From Fig. \ref{fig:int2} and Fig. \ref{fig:delay2}, we can find that
our proposed algorithm avoids congestion and minimizes delay at a cost of
a slight increase in the interference compared to the MA
algorithm. Our algorithm also provides robust routing results by
hierarchy level network routing, which forwards packets to the
nearest CPC base station.

\section{Conclusions}\label{sec:conclusion}
In this paper, we have introduced a dynamic and distributed routing
algorithm for multi-hop cognitive radio networks. The
routing algorithm minimizes the interference with the PUs from
multiple pairs of SU nodes to CPC base stations. We have used the
medial axis scheme with the relaxation factor as a reference path in the global procedure. 
In the local distributed procedure of the algorithm, we have used fictitious play learning to find a mixed-strategy  Nash equilibrium of  the dynamic routing game in the multi-hop CR network. We have shown that the
equilibrium can be found in closed form by backward induction. In simulations, we have observed that the mixed strategy of each SU node using
the learning algorithm converges within short iterations. The
algorithm significantly reduces interference with PU
networks  and achieves low packet delay
from source nodes to CPC base stations.

\bibliographystyle{IEEE}

\begin{thebibliography}{10}

\bibitem{Han2008}   Z. Han and K. J. R. Liu, {\em Resource allocation for wireless networks: basics, techniques, and applications}, Cambridge University Press, UK, April, 2008.

\bibitem{Hossain2009} E. Hossain, D. Niyato, and Z. Han, {\em Dynamic Spectrum Access in Cognitive Radio Networks}, {Cambridge University Press}, UK, 2009.

\bibitem{Schaar2009}    H. Shiang and M. V. D. Schaar, ``Distributed resource management in multihop cognitive radio networks for delay-sensitive transmission," {\em IEEE Trans. on Vehicular Technology}, vol. 58, pp. 941-953, Feburuary 2009.

\bibitem{Evans2005} M. Buddhikot, P. Kolodzy, S. Miller, K. Ryan, and J. Evans, "DIMSUMNet: New directions in wireless networking using coordinated dynamic spectrum access", in Proc. {\em IEEE Int. Symp. on a World of Wireless, Mobile and Multimedia Networks}, Giardini Naxos, Italy, Jun. 2005.

\bibitem{Cordier2009} O. Sallent, J. Perez-Romero, R. Agusti, and P. Cordier, "Cognitive pilot channel enabling spectrum awareness", in Proc. {\em Int. Conf. on Communications}, Dresden, Germany, Jun. 2009.

\bibitem{Guipponi2007} J. Perez-Romero, O. Salient, R. Agusi, and L. Guipponi, "A novel on-demand cognitive radio pilot channel enabling dynamic spectrum allocation", in Proc. {\em IEEE Int. Symp. New Frontiers in Dynamic Spectrum Access Networks}, Dublin, Ireland, Apr. 2007.

\bibitem{Hou2008} Y. Shi and Y. T. Hou, "A distributed optimization algorithm for multi-hop cognitive radio networks", in Proc. {\em Int. Conference on Computer Communications} pp. 1292-1300, Phoenix, AZ, USA, Apr. 2008

\bibitem{Medley2009}    L. Ding, T. Melodia, S. Batalama, and M. J. Medley, `` ROSA: Distributed joint routing and dynamic spectrum allocation in cognitive radio ad hoc networks," in Proc. {\em ACM International Conference on Modeling, Analysis and Simulation of Wireless and Mobile Systems}, Canary Islands, Spain, Oct. 2009.

\bibitem{Han2009} W. Saad, Q. Zhu, T. Ba\c{s}ar, Z. Han and A. Hjorungnes, ``Hierarchical network formation games in the uplink of multi-hop wireless networks", in {\em Proc. of Globecom},  Honolulu, Hawaii,  Nov. 30 -- Dec.4, 2009.

\bibitem{Jiang2006} J. Bruck, J. Gao, and A. Jiang, ``MAP: Medial axis based geometric routing in sensor networks", {\em Wireless Networks}, vol. 13, pp. 835-853, Oct. 2006.

\bibitem{Chen1997} W. C. Chan, T. -C. Lu, and R. -J. Chen, ``Pollarczek-khinchin formula for the M/G/1 queue in discrete time with vacations,'' {\em IEE Proc. Computers and Digital Techniques}, vol. 144, pp. 222-226, Jul. 1997.

\bibitem{basar95} T. Ba\c{s}ar  and G. J. Olsder,  ``Dynamic Noncooperative Game Theory,''  SIAM Series in Classics in Applied Mathematics, 2nd Edition, 1999.

\bibitem{Fudenberg98} D. Fudenberg  and D. K. Levine, {\em The Theory of Learning in Games},  {The MIT Press}, 1998.

\end{thebibliography}

\end{document}